\documentclass[12pt]{iopart}
\usepackage{iopams}  
\usepackage[latin1]{inputenc}
\usepackage[T1]{fontenc}
\usepackage{pifont}
\usepackage{textcomp}
\usepackage{mathptmx}
\usepackage[english]{babel} %
\usepackage{graphicx}
\begin{document}
\title{Laser cooling of a magnetically guided ultra cold atom beam  }
\author{A Aghajani-Talesh, M Falkenau, VV Volchkov, LE Trafford, T Pfau and A Griesmaier }
\address{Universität Stuttgart, 5.~Physikalisches Institut, Pfaffenwaldring 57\\ D-70550 Stuttgart, Germany}
\ead{a.aghajani@physik.uni-stuttgart.de, t.pfau@physik.uni-stuttgart.de}

\vspace{2pc}



\begin{abstract}
We report on the transverse laser cooling of a magnetically guided beam of ultra cold chromium atoms. Radial compression by a tapering of the guide is employed to adiabatically heat the beam. Inside the tapered section 
heat is extracted from the atom beam by a two-dimensional optical molasses perpendicular to it, resulting in a significant increase of atomic phase space density. A magnetic offset field is applied to prevent optical pumping to untrapped states. Our results demonstrate that by a suitable choice of the magnetic offset field, the cooling beam intensity and detuning, atom losses and longitudinal heating can be avoided. Final temperatures below 65~\textmu K have been achieved, corresponding to an increase of phase space density in the guided beam by more than a factor of 30.
\end{abstract}

\section{Introduction}

Magnetic guides loaded with ultra cold atoms from a magneto-optical trap (MOT) have in recent years been successfully established as intense sources of ultra cold atom beams \cite{Teo01,Cren02}. Maximum loading rates of up to $7\times10^9$ atoms s$^{-1}$at temperatures  below 400~\textmu K have been demonstrated \cite{Lahaye04}. 
For chromium, our group has recently reported the continuous loading of over $10^9$ atoms s$^{-1}$, albeit at somewhat higher temperatures than in \cite{Lahaye04}, by operating a moving molasses MOT in the field of a magnetic guide \cite{Griesmaier09}. 
Considerable effort has been invested in concepts for the exploitation of magnetic guides as ultra cold atom sources for the production of Bose-Einstein-Condensates (BECs) \cite{Mandonnet00}. 
A main driving force behind these efforts has been the prospective realization of a truly continuous atom laser \cite{SPREEUW95}, the matter wave analogue to a cw optical laser, which so far has only been achieved in pulsed or quasi-continuous mode  \cite{Anderson98,Mewes97,Bloch99,Chikkatur02,Hagley99,Cennini03,Robins08}. 
Two conceptually different approaches towards a coherent and continuous source of quantum matter have been proposed. The first approach is based on the direct condensation of a guided atom beam via evaporative cooling inside a magnetic guide \cite{Mandonnet00,Lahaye04}. The second approach employs a constantly refilled reservoir of ultra cold atoms from which atoms are transferred into a continuously maintained condensate \cite{Robins08,Chikkatur02}. In a recent proposal we have suggested a novel method for the loading of an optical dipole trap (ODT) from a magnetically guided atom beam \cite{Anoush09}. Our method is continuous in nature and can in principle be employed to operate the ODT as a pump reservoir for a pumped continuous atom laser. In the proposal we have identified the radial beam temperature $T_\mathrm{r}$ as the most critical parameter to make the transfer of atoms from the beam to the ODT more efficient. A reduction of the radial beam temperature is therefore the key to maximizing the potential pump rate for an atom laser that is fed from the dipole trap. 

In this article we present the transverse cooling of a magnetically guided beam of ultra cold chromium atoms using a two-dimensional optical molasses.
The cooling method we propose is akin to the Doppler-cooling of polarized atoms in a magnetic trap, which has been demonstrated for a number of elements, including chromium, calcium and neon \cite{Helmerson92,Setija93,Schreck01,Schmidt03,Spoden05,Hansen06}.
It differs in three respects from standard Doppler cooling of a free atom beam. First, the presence of a guide field splits the ground state into different Zeeman sub-states, which entails potential atom losses due to optical pumping from magnetically trapped into untrapped sub-states by the optical molasses beam. In order to suppress undesired optical pumping we therefore apply an additional magnetic offset field in the molasses region. Second, the beam that we intend to cool is already near the Doppler cooling limit. In order to enable effective cooling of the beam we use adiabatic compression by a tapering of the guide to increase the radial beam temperature before the beam enters the optical molasses. Third, the cooling is conducted on a beam that is magnetically confined in the radial direction, with the confining potential being affected by the magnetic offset field. 
In the following, we describe and discuss the implementation of our cooling method, its application to the atom beam and the resulting implications for the loading of a dipole trap from the beam.

\section{Experimental methods}
\begin{figure}
\begin{center}
\includegraphics[ height = 7 cm]{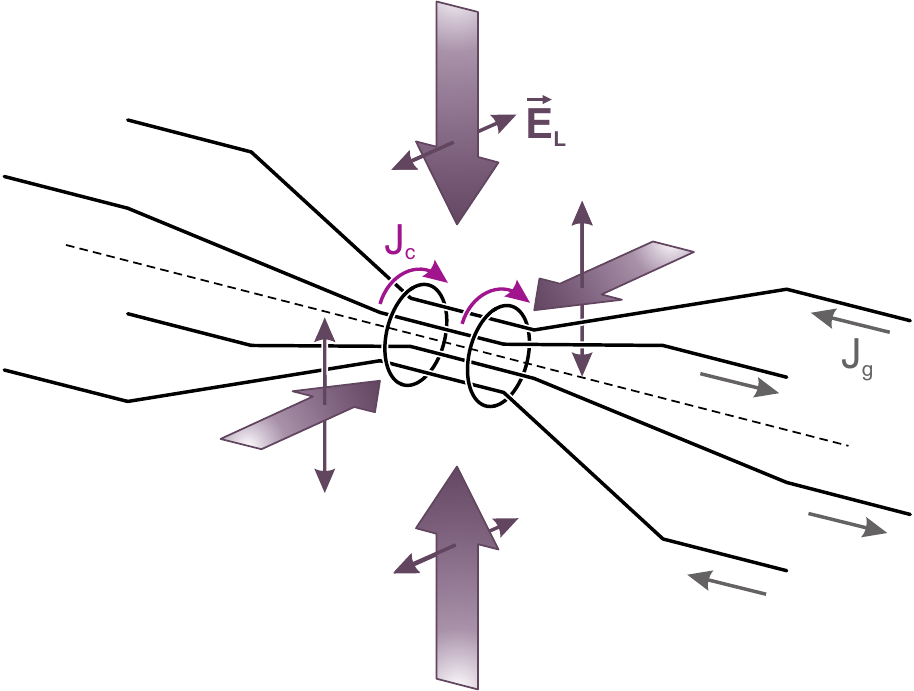}
\end{center}
\caption{Schematic illustration (unscaled) of the optical molasses employed for transverse cooling of the atom beam in a tapered section of the magnetic atom beam guide. The current $J_\mathrm{g}$, which flows in alternating opposing direction through the guide bars, generates a confining magnetic field, in which atoms are guided along central axis of the guide (dotted line). The tapering decreases the spacing between the guide bars from 46~mm to 9~mm, which results in compression and adiabatic heating of the passing atom beam. A two-dimensional optical molasses is formed by two pairs of counter propagating laser beams, which are indicated by flat arrows. The beams are linearly polarized in a lin~||~lin configuration with polarization perpendicular to the axis of the guide. 
The optical molasses operates on the $^{52}$Cr MOT transition at a wavelength of 426~nm.  
A pair of coils in the centre of the tapered section, carrying a current $J_\mathrm{c}$, provides a magnetic offset field in order to suppress optical pumping to untrapped states in the optical molasses.  
}
\label{fig:TapSec}
\end{figure}

\begin{figure}
\begin{center}
\includegraphics[ height = 7 cm]{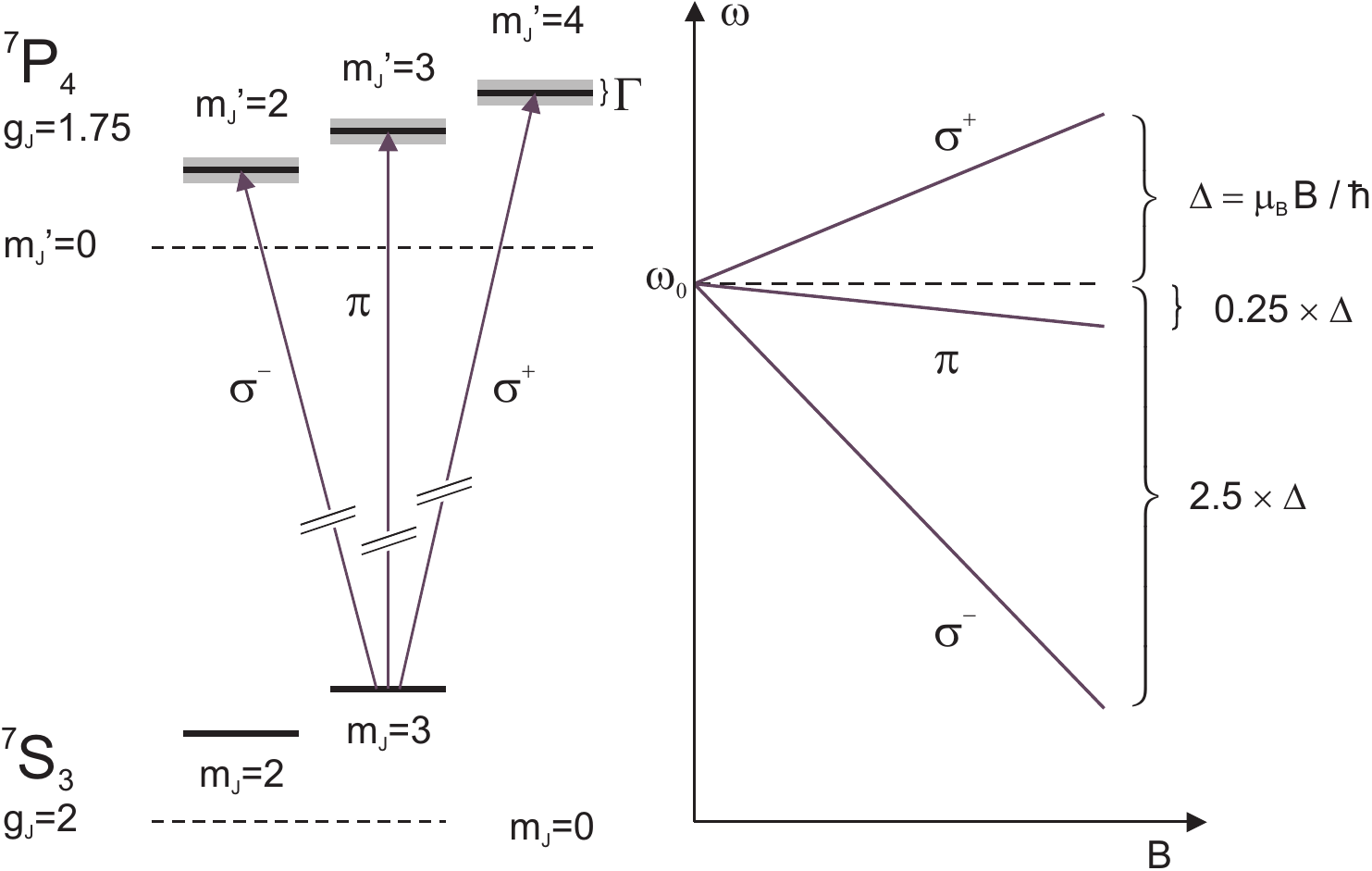}
\end{center}
\caption{Selected Zeeman line shifts induced by an offset field B$_\mathrm{0}$ for the $^{52}$Cr $^7\mathrm{S}_3\,\rightarrow\, ^7\mathrm{P}_4$ transition, which used for the transverse cooling of the atom beam.  The $\sigma^+$ transition from the $m_j=+3$ to the $m'_j=+4$ experiences a positive shift $\Delta= \mu_\mathrm{B} B_\mathrm{0} \hbar^{-1}$, which amounts to 0.28~$\Gamma \mathrm{G}^{-1}$. The corresponding $\sigma^-$ and $\pi$ transitions, on the contrary, experience negative shifts of $-2.5 \Delta$ and $-0.25 \Delta$, respectively. This allows to perform Doppler cooling on the closed $\sigma^+$ transition exclusively by applying a sufficiently strong offset field B$_\mathrm{0}$.
}
\label{fig:Zeeman}
\end{figure}

The experiments described in this article were conducted using the guided chromium atom beam apparatus described in \cite{Griesmaier09,GreinerA07}. Here, only its main features are summarized briefly. The atom beam apparatus comprises a linear quadrupole magnetic guide that is constituted by four rectangularly arranged copper bars with spacing $d$. The bars carry a current $J_\mathrm{g}$ of typically 180~A in alternating opposing directions. The resulting magnetic field provides two-dimensional radial confinement via its interaction with the magnetic moments of the guided atoms. 
Adopting the central axis of the guide as z-axis of a cartesian coordinate system, we can describe the total guiding field by $\left(b_\mathrm{g} \, x, \, b_\mathrm{g} \, y, \, B_\mathrm{0}\right)$. This description includes a potentially non-zero z-component, in order to consider the case of an additionally applied longitudinal uniform magnetic offset field with magnitude $B_\mathrm{0}$. The variable $b_\mathrm{g}$ represents the absolute value of the magnetic field gradient, which, with $\mu_\mathrm{0}$ denoting the vacuum permeability, is given by
\begin{equation}
b_\mathrm{g} = \frac{4 J_\mathrm{g} \mu_\mathrm{0}}{d^{2} \pi}.
\label{eq:MagGrad}
\end{equation}  
For an atom in a state with magnetic quantum number $m_J$ and Landé-factor $g_J$ the confining potential 
$U_\mathrm{g}$ of the guide as a function of the radial distance $r$ from the guide axis can be written as
\begin{equation}
U_\mathrm{g}(r)=  m_J g_J \mu_\mathrm{B} \, \sqrt{b_\mathrm{g}^2 \, r^2 + B_\mathrm{0}^2}.
\label{eq:GuidePot}
\end{equation}  
Here $\mu_\mathrm{B}$ denotes the Bohr magneton. Equation \ref{eq:MagGrad} and \ref{eq:GuidePot} relate the strength of the confining potential with the spacing of the guide bars.

The magnetic guide is loaded with atoms from a moving molasses MOT (MMMOT) that is operated directly in the guide field at a bar spacing of 46~mm, which corresponds to magnetic field gradient of 13.5~G\,cm$^{-1}$. The MMMOT injects a continuous beam of ultra cold $^{52}$Cr ground state atoms into the guide, with a mean beam velocity $v_\mathrm{b} $ tuneable from about 1\,ms$^{-1}$ to over 10~ms$^{-1}$. Radial and axial beam temperatures as low as 180~\textmu K can be achieved, which is near the Doppler limit of 125~\textmu K for chromium. 

We conduct the subsequent transverse cooling of the emitted atom beam by employing an optical molasses inside a tapered section of the guide, as illustrated by figure \ref{fig:TapSec}. The tapering is located at a distance of 0.75~m away from the MMMOT. It reduces the bar spacing from 46~mm to 9~mm, thus gradually increasing the magnetic field gradient over a distance of 100~mm from initially 13.5~G\,cm$^{-1}$ to 355~G\,cm$^{-1}$. The tapered section has a length of 50~mm, after which the spacing of the bars re-expands until the field gradient again reaches its original value of 13.5~G\,cm$^{-1}$.   
The purpose of the tapered section is to compress and, thereby, to adiabatically heat the beam before it enters the optical molasses. The optical molasses is oriented perpendicularly to the atom beam. It is generated by two pairs of retro-reflected laser beams, which both are linearly polarized in a lin~||~lin configuration with polarization perpendicular to the atom beam axis, as shown in figure \ref{fig:TapSec}. 
The beam intensities are, within the limits of a retro reflected configuration, approximately equal. Throughout this article we will usually quote the peak intensity $I_\mathrm{L}$ of a single molasses beam.
The molasses beams have identical elliptical beam profiles with respective beam waists of 13.1~mm along the guide axis and 2.5~mm perpendicular to it.
 An axially elongated beam profile has been chosen in order to maximize the distance over which the atom beam is illuminated, and thus the number of photons that can be scattered by an atom for a given beam intensity during its passage through the molasses beam. 

The optical molasses is operated on the $^{52}$Cr $^7\mathrm{S}_3\,\rightarrow\, ^7\mathrm{P}_4$ transition at 426~nm, which is a practically closed transition that is also employed for the MMMOT. It has a line width $\Gamma$ of $2\pi \times 5$~MHz and a saturation intensity $I_\mathrm{s}$ of 8.5~mW\,cm$^{-2}$. From equation~\ref{eq:GuidePot} it can be seen that only atoms with $m_J > 0$ can be confined by the magnetic field of the atom guided. 
In order to prevent atom losses in the optical molasses, it would therefore be desirable to scatter predominantly $\sigma^+$-polarized photons. 
Due to the presence of the guide field, it is, however, not possible to prepare the molasses beam in a purely $\sigma^+$-polarized state. 
Instead, we can employ a magnetic offset field $B_\mathrm{0}$ to split the Zeeman lines far enough apart such they can be selectively addressed. Figure \ref{fig:Zeeman} illustrates the Zeeman line shift of transitions from the $m_J=+3$ ground state.
Due to a lower Landé factor of $g_J=1.75$ in the upper level, the $\sigma^+$-transition experiences a positive shift towards higher frequencies, whereas the $\sigma^-$ and $\pi$ transitions experience negative shifts. 
Provided that the splitting is sufficiently strong it is thus possible to tune the molasses beam near-resonant with the $\sigma^+$ and off-resonant with both $\sigma^-$ and $\pi$ transitions.     

The offset field is produced by a coaxial pair of identical copper coils that carry the current $J_\mathrm{c}$ with equal orientation. The coils are positioned concentrically with the beam axis around the centre of the tapered region, as shown in figure \ref{fig:TapSec}. The coils have an inner diameter of 42~mm and an inner spacing of 22~mm at a coil length of 15~mm. The coils thus approximate a Helmholtz coil pair configuration. We therefore regard the offset field experienced by the atoms in the optical molasses to be uniform and perpendicular to the guide field.

\section{Experimental observations}

\begin{figure}
\begin{center}
\begin{minipage}{0.46 \textwidth} (a) \medskip \\ \includegraphics[width=\textwidth]{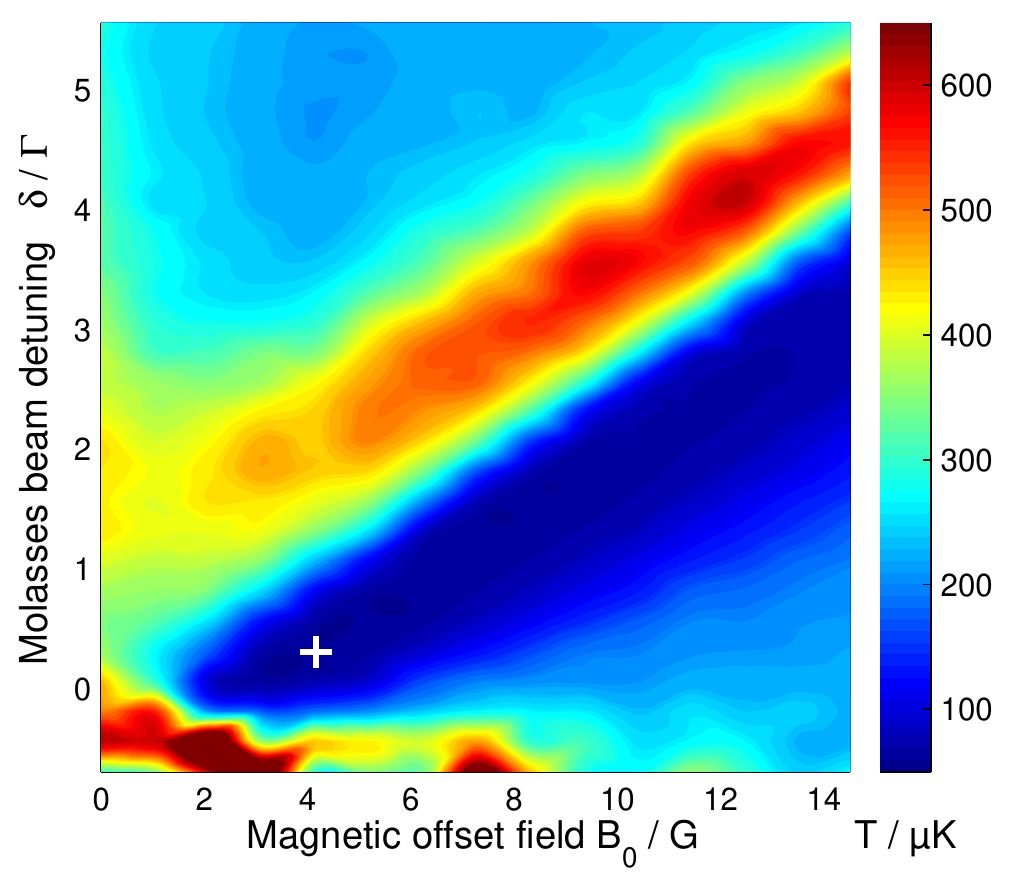} \end{minipage} 
\begin{minipage}{0.46 \textwidth} (b) \medskip \\ \includegraphics[width=\textwidth]{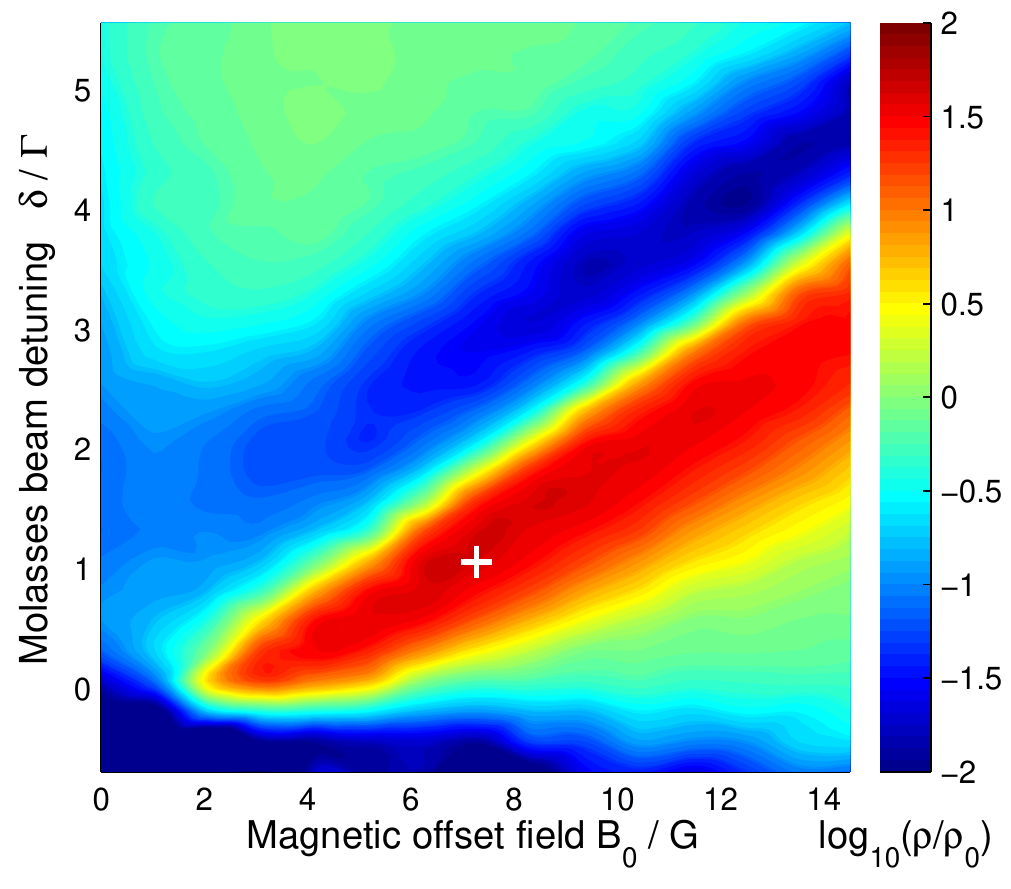} \end{minipage}
\end{center}
\caption{(a) Colour coded contour plot of the radial temperature of the decompressed beam after passing the tapered optical molasses section.  The temperature has been recorded as function of the magnetic offset field and of molasses beam the detuning at a beam velocity of 5.2~m\,s$^{-1}$ and a peak molasses beam intensity of 0.3 $I_\mathrm{s}$. The white cross indicates the minimal radial temperature of 62~\textmu K, which has been observed for an initial temperature of 230~\textmu K. (b) Associated gain in phase space density ($\rho / \rho_\mathrm{0}$) obtained by including atom flux and axial temperature data. A maximum gain of more than a factor 40 can be observed (white cross). The minimum temperature value on the left corresponds to a gain of a factor 30.}
\label{fig:PSDTemp}
\end{figure}

\begin{figure}
\begin{center}
\includegraphics[width=0.75\textwidth]{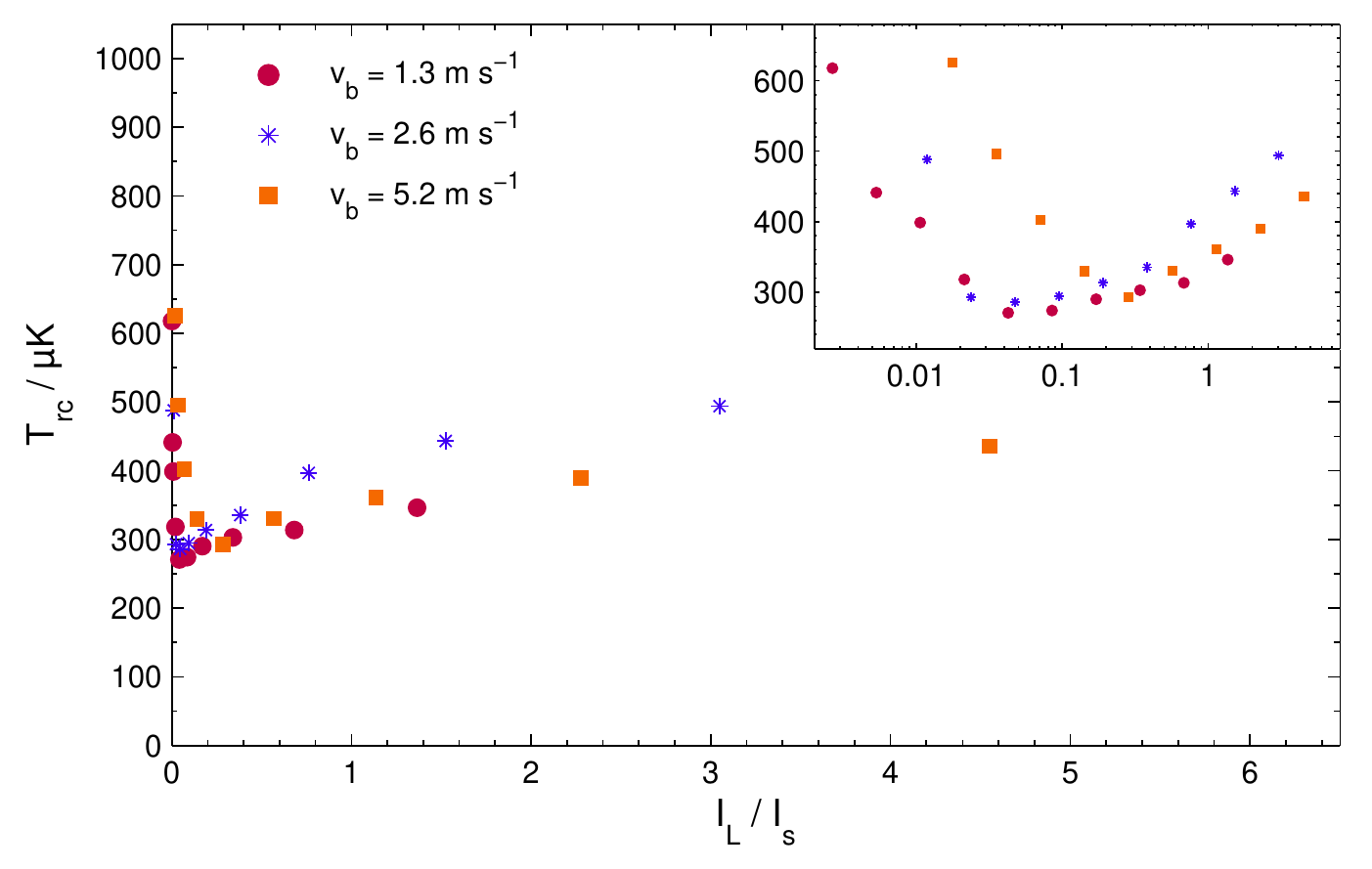}
\end{center}
\caption{Temperature $T_\mathrm{rc}$ of the compressed atom beam over molasses beam intensity for three distinct beam velocities $v_\mathrm{b}$. The inset displays the same data with a logarithmic intensity axes. For low intensities $I_\mathrm{L} \ll I_\mathrm{s}$ the beam temperature is velocity dependent and has thus not reached a steady-state during the time of passage through the optical molasses. For $I_\mathrm{L} \geq I_\mathrm{s}$ a linear increase of the final temperature due to saturation effects can be observed. The minimum observed temperature is more than two times higher than the Doppler cooling limit of an unconfined Cr atom cloud. }
\label{fig:TempVsIntensity}
\end{figure}

Our studies of the cooling with the optical molasses are based on the characterisation of the beam in terms of the total atom flux, the radial and axial beam temperature. The methods that we have adopted for the measurement of these beam properties are identical to the methods described in \cite{Griesmaier09}. They rely on the illumination of the beam with a near resonant probe beam of known intensity and the imaging of the scattered photons with a calibrated CCD-Camera. Since the four optical molasses beams occupy a large fraction of the optical access to the tapered section of the guide, we have performed our measurement on the decompressed beam at a succeeding section of the guide, where the magnetic guiding potential is equal to the uncompressed initial potential. 

Figure \ref{fig:PSDTemp}\,(a) presents the results from a measurement of the radial temperature as a function of the offset field $B_\mathrm{0}$ and the molasses detuning $\delta$. The beam has an initial radial temperature of 230~\textmu K at a beam velocity of 5.2~m\,s$^{-1}$. The intensity of the molasses beam was set to 0.3~$I_\mathrm{s}$. The observed temperatures are represented by a colour coded temperature map that has been obtained from interpolation between individual measurements. It can be seen that effective cooling below the initial temperature can only be achieved inside a wedge shaped region of the plot. The slopes of its borders closely match the respective Zeeman shifts of the $\sigma^+$ and $\pi$ transition from the $m_J=+3$ state, which are given in figure \ref{fig:Zeeman}. This indicates that the magnetic offset field plays a vital role in the cooling process. Without magnetic field, the Zeeman-levels are degenerate and no cooling can be observed. 
At a field of about 4.1~G an optimal value $\delta_\mathrm{min}$ of the detuning can be found that yields a minimum final temperature $T_\mathrm{min}$ of 62~\textmu K.
At this detuning the molasses is sligthly red-detuned relative to the $\sigma^+$ transition ($\approx -1~\Gamma$) \footnote{It should be noted here, that, while we can control the molasses beam frequency with a precision much better than the atomic line width $\Gamma$, we can only achieve an absolute frequency accuracy of about 0.5~$\Gamma$.}.
For higher values of $\delta$, the molasses beam becomes blue-detuned, which results in heating of the atom beam. Towards lower detunings, the beam becomes eventually near-resonant and blue-detuned with the $\pi$ transition, which again leads to heating. When the magnetic field is increased further above 4.1~G we observe that the cooling becomes less efficient, resulting in higher final temperatures. This behaviour is caused directly by offset field, as it reduces the radial compression in the tapered section of the guide. This reduces the cooling due to the decompression after the optical molasses. 

For each data point in figure \ref{fig:PSDTemp}\,(a) we have also recorded the corresponding atom flux$ \phi$ and axial temperature $T_\mathrm{a}$, which allows us to deduce the associated phase space density $\rho$ of the beam according to \cite{Griesmaier09}
\begin{equation}
\rho=\frac{1}{2\pi}\frac{\phi}{v_\mathrm{b}}\left( \frac{\mu b_\mathrm{g}}{k_\mathrm{B} T_\mathrm{r}} \right)^2 \left( \frac{2\pi \hbar^2}{m\, k_\mathrm{B} \left(T_\mathrm{r}^2 T_\mathrm{a}\right)^{1/3}}\right)^{3/2}.
\end{equation}
Here $m$ denotes the atomic mass, $\mu$ the atomic magnetic moment and $k_\mathrm{B}$ the Boltzmann constant. In figure \ref{fig:PSDTemp}\,(a) the gain in phase space density compared to its initial value $\rho_\mathrm{0}$ is presented. The gain in phase space density coincides well with the cooling of the beam. For the minimum temperature in figure \ref{fig:PSDTemp}\,(a) we observe a gain of more than a factor 30, which yields a final phase space density of $3 \times 10^{-8}$. 

Beside the molasses beam detuning and the offset field, the intensity $I_\mathrm{L}$ of the molasses beam is expected to have a significant influence on the cooling process. For low intensities, the duration of the passage through the optical molasses might be insufficient for the beam to reach a steady state temperature, which should then manifest itself as a strong intensity dependence of the final temperature. High intensities with $I_\mathrm{L} \geq I_\mathrm{s}$ might lead to a specific intensity dependence due to saturation effects \cite{Lett89}. 
In figure \ref{fig:TempVsIntensity} we present the results from an investigation of the intensity dependence of the radial temperature at three different beam velocities. We conducted our measurements at a fixed magnetic offset field of 23.4~G. The molasses beams had a detuning relative to the Zeeman shifted $\sigma^+$ line of $-1.25\,\Gamma$, which for the chosen offset field corresponds to a detuning of about $+7\,\Gamma$ to the $\pi$ transition. Note that  
the temperature values given in figure \ref{fig:TempVsIntensity} refer, in contrast to the values given in figure \ref{fig:PSDTemp}, to the temperature $T_\mathrm{rc}$ of the compressed beam inside the tapered section of the guide. They have been deduced from the measured uncompressed temperature values via a conversion that is based on numerical simulations of the decompression process. For this purpose we have calculated a large number of atom trajectories with initial values randomly sampled according to a thermal distribution \cite{Anoush09}. From figure \ref{fig:TempVsIntensity} it can be seen that for the lowest intensities in the plot, the final temperature decreases with increasing intensity. In addition, lower beam velocities, which are equivalent to a higher number of scattered photons, lead, for the same intensity, to lower final temperatures. Both observations combined provide firm evidence that in the case of the two lowest velocities for $I_\mathrm{l}\lesssim 0.04\,I_\mathrm{s}$ and in the case of $v_\mathrm{b} = 5.2$~ms$^{-1}$ for $I_\mathrm{l}\lesssim 0.2\,I_\mathrm{s}$ the atom beam does not reach a steady state. 
When the intensity is increased above these values, we find that the final temperatures do not further decrease. At the same time, we observe that the temperature curves for all regarded beam velocities begin to coincide. We interpret this as a strong indication that the beam has entered a steady state regime. 
Finally, for intensities exceeding $0.5\,I_\mathrm{s}$, the radial temperature increases approximately linearly with the beam intensity, which we attribute to the beginning saturation of the cooling transition \cite{Lett89}. 
The minimum final temperature of about 270~\textmu K is considerably higher than the Doppler limit of 125~\textmu K, that would be expected for an optical molasses. We suppose that this might be due to the presence of the strong magnetic field gradient in the tapered section, which is more than 25 times higher than the typical magnetic field gradients in a chromium MOT. A comprehensive theoretical description of the cooling process that accounts for the magnetic field is, however, beyond the scope of this article. 

\section{Conclusion}
A method for the transverse cooling of a magnetically guided atom beam has been presented. We have demonstrated that by magnetic compression followed by transverse Doppler cooling in an optical molasses and subsequent decompression can be employed to increase the phase space density of an ultra cold atom beam resulting in final temperatures below the Doppler cooling limit. Our method could in principle be also applied to other atomic species than chromium. It is especially well suited to rare earth elements with large magnetic moments and similar optical properties such as erbium and dysprosium, which have recently gained much attention in conjunction with ultra cold atom experiments \cite{McClelland06,Lu09}.
Our results are directly applicable to the loading of an optical dipole trap as suggested in \cite{Anoush09}. For an experimentally realized beam with a final radial temperature of 63~\textmu K, a beam velocity of 1.2~ms$^{-1}$ and a longitudinal temperature of 250~\textmu K, we estimate based on \cite{Anoush09} a loading efficiency of more than 5~\%, corresponding to a loading rate of over 10$^7$ atoms\,s$^{-1}$. Compared to the typical production rate of chromium atoms in Cr-BEC \cite{Griesmaier05}, which is $3\times10^3$ atoms s$^{-1}$ this would provide an excellent starting point for a significant improvement of the Cr-BEC production rate.

\ack
Financial support by the Landes\-stif\-tung Ba\-den-Wür\-ttem\-berg under contract No.~0904Atom08 is gratefully acknowledged by the authors of this article. AA-T acknowledges financial support by the Stu\-dien\-stif\-tung des Deutschen Volkes. \bigskip

\vspace{0 cm}
\bibliographystyle{unsrt}
\bibliography{DC-Article-IOPStyle}

\end{document}